\documentclass[11pt,twoside]{article}

\usepackage{asp2006}
\usepackage{epsf}
\usepackage{psfig}
\usepackage{lscape}

\begin{document}
\title{Star Formation and the Interstellar Medium in Nearby Tidal Streams (SAINTS)}   %%% Fill in title
\author{S. J. U. Higdon,\altaffilmark{1} J. L. Higdon,\altaffilmark{1} B. J. Smith,\altaffilmark{2}
M. Hancock,\altaffilmark{3} and C. Struck\altaffilmark{4}}  

\altaffiltext{1}{Georgia Southern University, Department of Physics, Statesboro, GA 30458}
\altaffiltext{2}{East Tennessee State University, Department of Physics and Astronomy, Johnson City, TN 37614}
\altaffiltext{3}{University of California, Riverside, Riverside, CA 92521}
\altaffiltext{4}{Iowa State University, Department of Physics and Astronomy, Ames, IA 50011} 

\begin{abstract} %%% Abstract

We compare Spitzer Infrared Spectrograph observations of SQ-A \& SQ-B
in Stephan's Quintet, Ambartzumian's knot in Arp 105, Arp 242-N3, Arp
87-N1, a bridge star forming region, NGC 5291~N and NGC 5291~S. The
PAHs tend to be mainly neutral grains with a typical size of 50 - 100
carbon atoms. The interstellar radiation field is harder than typical
starburst galaxies, being similar to that found in dwarf galaxies. The
neon line ratios are consistent with a recent episode of star
formation $\la$ 5 million years ago.  We detect emission from $\sim$
10$^6$ M$_{\odot}$ of warm H$_2$ in SQ-A and Arp~87N1 and $\sim$
10$^5$ M$_{\odot}$ in SQ-B. These results are similar to those derived
for the tidal dwarf galaxies (TDGs) NGC 5291 N and NGC 5291 S and are
consistent with emission from photodissociation regions.  Using our 8
$\mu$m images of 14 interacting systems we identify 62 tidal star
forming knots (TSFKs). The estimated stellar masses range from super star
cluster (10$^4-10^6$M$_{\odot}$) to TDG ($\sim10^9$M$_{\odot}$)
sizes. The median stellar mass is 10$^8$ M$_{\odot}$. The stellar
mass, with some scatter, scales with the 8 $\mu$m luminosity and tends
to be an order of magnitude smaller than the KISS sample of star
forming dwarfs. An exception to this are the TSFKs in Arp 242 which
have stellar masses similar to the KISS dwarfs. The TSFKs have
``notched'' 3.6 - 8 $\mu$m spectral energy distributions (SEDs)
characteristic of star forming regions. The TSFKs, form two distinct
clumps in a mid-infrared color diagram. There are 38 red-TSFKs with
$[4.5] $ - $[8.0] $ $>$ 3 and $[3.6] $ - $[4.5] $ $<$ 0.4. This
populations has significantly enhanced non-stellar emission, most
likely due to PAHs and/or hot dust, relative to normal spirals and the
KISS sample of dwarfs.  The second group of 21 sources has 1.2 $<$
$[4.5] $ - $[8.0] $ $<$ 3 and $[3.6] $ - $[4.5] $ $<$ 0.4. This
population overlaps with the colors of star forming dwarf and spiral
galaxies.  The redder $[4.5] $ - $[8.0] $ population tends to have the
sources with a rising 8-24 $\mu$m SED while the blue population tends
to contain the sources with a descending SED. The rising SED is
typical of spiral and starburst galaxies with a dominant 40 $-$ 60 K
dust component and the declining SED probably indicates a dominant hot
dust component.
\end{abstract}

\section{Introduction}

In addition to triggering starbursts and active galactic nuclei,
mergers of dusty, gas rich disk galaxies frequently lead to the
formation of tidal tails that can stretch many disk diameters from the
site of the collision (\citealp{toomre72}; \citealp{schweizer78};
\citealp{sanders96}). These structures tend to be HI rich with blue
optical colors, reflecting both their origin in the outer spiral disk
and on-going star formation (\citealp{van79}; \citealp{schombert90};
\citealp{mirabel91}; \citealp{hibbard96}).  \citet{zwicky56} proposed
that dwarf galaxies might form out of self-gravitating clumps within
tidal tails, and indeed, concentrations of gas and star forming
regions are commonly found there, ranging in size from super star
clusters (SSCs, 10$^4-10^6$M$_{\odot}$) to tidal dwarf galaxies (TDGs,
$\sim10^9$M$_{\odot}$).

Tidal star forming knots (TSFKs) and the formation of gravitationally
bound TDGs, formed either via tidal interactions between the parent
galaxies or from ram-sweeping of debris material, may play an
important role in galaxy formation and evolution.  \citet{higdon06a},
hereafter HHM06, found tidal bridges and tails associated with TDGs
and TSFKs in NGC 5291, indicating further tidal interaction amongst
the star forming knots. TDGs may be useful as local analogs of the
multiple mergers of small dwarf-like galaxies at high redshift.  More
importantly, dwarf galaxies are the most common galaxy type in the
current epoch, and TDGs may contribute significantly to this
population in some environments, for example, in compact groups
\citep{hunsberger96}.

Here we present some new results from our Spitzer study of Star
Formation and the Interstellar Medium in Nearby Tidal Streams
(SAINTS). We have selected 12 pre-merger binary pairs with prominent
optical tails and/or bridges (\citealp{smithb07};
\citealp{higdon08}). This is complemented by our study of two more
complex systems: NGC~5291 where ram-sweeping is occurring in addition
to a tidal interaction, and TSFKs in Stephan's Quintet. Using the
Spitzer 8 $\mu$m images we have identified 62 TSFKs. We discuss the
Infrared Spectrometer (IRS, \citealp{houck04}) observations of seven
bright TDG candidates along with the Infrared Array Camera (IRAC,
\citealp{fazio04}) and Multiband Imaging Photometer for Spitzer (MIPS,
\citealp{rieke04}) images of the full sample and address some key
questions concerning the nature of TDGs and star formation in tidal
streams.
 
\section{The Nature of Tidal Dwarf Galaxies and Tidal Star Formation}

Figure 1 shows the IRS Short-Low (IRS-SL) observations of NGC 5291~N
and NGC 5291~S (from HHM06), SQ-A \& SQ-B in Stephan's Quintet and
Ambartzumian's knot in Arp 105 \citep{higdon08}, and Arp 242-N3, and
Arp 87-N1, the bridge star forming region in Arp 87.  The TDGs/TSFKs
are rich in atomic and molecular emission features from the ISM,
including fine structure lines, e.g., [Ne\kern.2em{\sc ii}] 12.81
$\mu$m, [Ne\kern.2em{\sc iii}] 15.56 $\mu$m, polycyclic aromatic
hydrocarbons (PAHs) and warm H$_2$. The mid-infrared spectra of dwarf
galaxies are known to differ substantially from those of spirals, with
weaker PAH emission features and higher [Ne\kern.2em{\sc
    iii}]/[Ne\kern.2em{\sc ii}] ratios \citep{madden06}. However, it
is unknown whether the mid-infrared spectra of most TSFKs
more closely resemble those of spirals or dwarfs, since only a handful
have been studied in detail with Spitzer IRS spectra to date (e.g.,
HHM06; \citealp{higdon08}).  For the TDGs in NGC 5291,
the spectra more closely resemble those of star forming dwarfs than
spirals.

The data was analyzed using SMART \citep{higdon04} and the strengths
of the PAH features were measured using PAHFIT \citep{smith07}. In the
proceeding sections the results are used to address five questions.

\begin{figure*}[ht]
\begin{center}
\scalebox{0.5}{\rotatebox{0}{\includegraphics{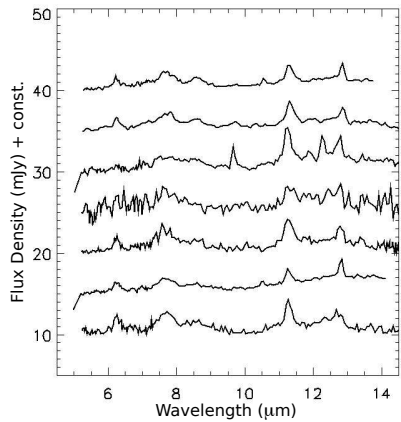}}}
\end{center}
\caption[]{IRS-SL Spectra in the rest-frame wavelength. From top to
  bottom, NGC 5291 N, NGC 5291 S, Stephan's Quintet-A, Stephan's
  Quintet-B, Arp 87-N1, Ambartzumian's Knot in Arp 105 \& Arp 242-N3.}
\end{figure*}

\subsection{Are the PAH Grains Large? Are they Charged ?}

\citet{draine01} model the ratio of the 6.2 to 7.7 $\mu$m features,
which are both from C-C stretching modes, to derive the PAH
ion$/$grain size. The number of carbon atoms decreases as the ratio
increases.  The 11.3 $\mu$m feature is from a C-H out of plane bending
mode and the ratio of the 11.3 to 7.7 $\mu$m band strength gives a
measure of the PAH ion fraction. The ratio of neutral to ionized PAHs
increases as the ratio increases. The PAHs in the TSFKs
tend to have a strong C-H out-of-plane bending mode resulting in a
mainly neutral population of PAH grains.  PAH$_{N}$ $\sim$ 80 - 90 \%
in Ambartzumian's knot, Arp 242-N3, SQ-B and Arp 87-N1 and around
PAH$_{N}$ $\sim$60 \% in the remainder of the sample. Arp 242-N3 and
Ambartzumian's knot in Arp~105 have small PAHs, N$_C$ $\sim$ 50. The
rest of the sample have larger PAHs, N$_C$ $\sim$100 (the model fit to
SQ-A is not well constrained).

\subsection{Are the Interstellar Radiation Fields (ISRF) in TSFKs and Dwarfs Similar?} 

The TSFKs in SQ-A and SQ-B have [S\kern.2em{\sc iv}]$/$[S\kern.2em{\sc
    iii}]~ and [Ne\kern.2em{\sc iii}]$/$[Ne\kern.2em{\sc ii}] line
ratios indicating moderate excitation consistent with our earlier
results for NGC 5291 N and NGC 5291 S (HHM06) and with the overlap
region in the Antennae \citep{verma03}. The ISRF in TSFKs is harder
than typical starburst galaxies, being similar to that found in dwarfs, but
softer than the ISRFs found in the extreme low metallicity Blue Compact Dwarfs
\citep{verma03}. The neon line ratios in the TSFKs are consistent with
the Starburst99 models in \citet{thornley00} for a recent episode of
star formation $\la$ 5 million years ago, i.e., on-going star
formation.

\subsection{How Much Warm H$_2$ do the TSFKs Contain?}
We detect emission from $\sim$ 10$^6$ M$_{\odot}$ of warm H$_2$ in
SQ-A and Arp~87-N1 and $\sim$ 10$^5$ M$_{\odot}$ in SQ-B. These results
are similar to those derived by HHM06 for NGC 5291 N and NGC 5291 S
(HHM06) The warm H$_{2}$ masses of $10^5 -10^6$M$_{\odot}$ are about
100 - 1000 times smaller than the average warm molecular mass of 2
$\times$ $10^8$ M$_{\odot}$ measured in a sample of 59 ULIRGs
\citep{higdon06b}, but the results for both galaxy types are consistent
with an origin in photodissociation regions (PDRs).

\begin{figure*}[ht]
\begin{center}
\scalebox{0.5}{\rotatebox{0}{\includegraphics{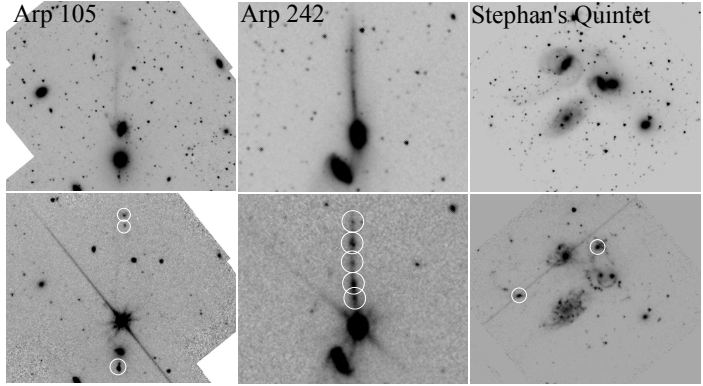}}}
\end{center}
\caption[]{IRAC images of (left-right) Arp 105, Arp 242 \& Stephan's
  Quintet. Top row 3.6 $\mu$m and bottom row 8 $\mu$m. TSFKs are
  identified with a white circle. North is top and East is left. A square root
  transform is used.}
\end{figure*}

\subsection{Can the Mid-infrared Colors and SEDs Distinguish TSFKs from other Galaxy Types?}

Figure 2 shows three examples of the 3.6 $\mu$m and 8 $\mu$m
images for Arp 105, Arp 242 (The Mice) and Stephan's Quintet. The
locations of the TSFKs are indicated with white circles on the
figures. 95\% of the sources in the 14 systems have a ``notched'' IRAC
SED characteristic of star forming regions. 79\% have 24 $\mu$m
detections or meaningful upper limits. These are classified into four
groups depending on the 8 - 24 $\mu$m slope. There are 18 rising
(f$_{24} -f_8 > 0.3\times f_{24}$), 12 declining SED (f$_{8} -f_{24} >
0.3\times f_{24}$), 9 flat/slowly rising (f$_{24} >f_8 $ and f$_{24}
-f_8 < 0.3\times f_{24}$), and 8 flat/slowly declining (f$_{24} < f_8 $
and f$_{8} -f_{24} < 0.3\times f_{24}$). The rising SED is typical of
spiral and starburst galaxies with a dominant 40 $-$ 60 K dust
component and the declining SED probably indicates a dominant hot dust
component. Far-infrared observations are needed to confirm this
interpretation.

\begin{figure*}[ht]
\begin{center}
\scalebox{.55}{\rotatebox{0}{\includegraphics{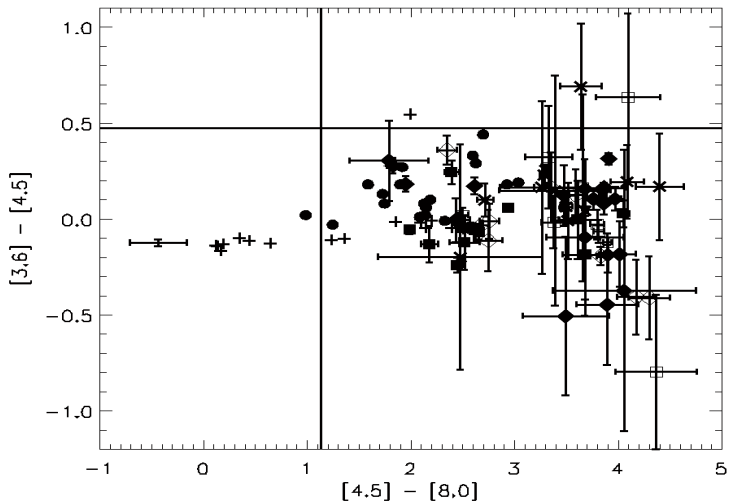}}}
\end{center}
\caption[]{The horizontal and vertical lines mark the color
  zero-points (i.e. where the two flux densities are equal). The lower
  right quadrant contains sources with active star formation. The
  TSFKs are coded according to 8 - 24 $\mu$m SED shape: Rising -
  filled diamond, Flat/Rising - diamond; Descending - solid square;
  Flat/Descending - square; unknown - X. For comparison the
  \citet{pahre04} galaxy sample are shown as crosses and the
  \citet{rosenberg06} star forming dwarfs are shown as solid circles.
}
\end{figure*}

Figure 3 shows that TSFKs form two distinct clumps on an IRAC 2-color
diagram. The first group has 38 red-TSFKs with $[4.5] $ - $[8.0] $ $>$
3 and $[3.6] $ - $[4.5] $ $<$ 0.4. This populations has significantly
enhanced non-stellar emission, most likely due to PAHs and/or hot
dust, relative to normal spirals \citep{pahre04} and star forming
dwarf galaxies (only overlaps with one dwarf in the KISS sample,
\citealp{rosenberg06}). The second group of 21 sources has 1.2 $<$
$[4.5] $ - $[8.0] $ $<$ 3 and $[3.6] $ - $[4.5] $ $<$ 0.4. This
population overlaps with the majority of the KISS dwarfs and the
spiral galaxies from Pahre's sample.  The redder $[4.5] $ - $[8.0] $
population tends to have the sources with a rising 8-24 $\mu$m SED
while the blue population tends to contain the sources with a
descending SED.

\subsection{What is the Range in Stellar Masses? Do the Stellar Masses Depend on the SED Type or the 8 $\mu$m Luminosity?} 

Preliminary stellar masses are estimated using the 4.5 $\mu$m data
(e.g., \citealp{oh08}) and assuming minimal non-stellar emission in
the band. We find a median M$_*$ = 10$^8$ M$_{\odot}$, with a wide
range: 4.0 $\times$ 10$^6$ to 1.3 $\times$ 10$^9$ M$_{\odot}$. The 18
rising sources have a median stellar mass of 1.0 $\times$ 10$^8$
M$_{\odot}$. The 12 declining sources have a higher median stellar
mass of 4.0 $\times$ 10$^8$ M$_{\odot}$. Half of the declining sources
are in Arp 242, which probably biases the median stellar mass to be
greater than the average for the whole sample. The sources with
flat or slowly rising/declining SEDs have stellar masses similar to
that of the rising sources.
 
The stellar mass, with some scatter, tends to scale with the 8 $\mu$m
luminosity. For a given 8 $\mu$m luminosity the KISS dwarfs have a more
massive stellar component.  The median stellar mass for the KISS dwarfs
is 1.0 $\times$ 10$^9$ M$_{\odot}$, which is a factor of ten larger
than the median value for the TSFKs. 6/8 TSFKs in Arp 242 have
declining SEDs. Four of these TSFKs overlap with 3 of the dwarfs from the
KISS sample.

\section{Future Work}

Studies of TDGs and smaller knots of star formation in tidal features
help us better understand the dwarf galaxy population as a whole. The
mid-infrared emission from the ISM in TSFKs are rich in
atomic, molecular and PAH emission features. The study of these
systems is still in its infancy as few have been studied in
detail. The next generation of telescopes, for example Herschel, JWST
and ALMA offer the promise of studying both fainter objects and larger
samples. In particular, such studies will provide information about what
fraction are truly primordial building blocks of massive galaxies left
over from an early epoch of galaxy formation, and what fraction may
have been born in tidal interactions. In time we will perform a more
comprehensive census of star formation and the interstellar medium in
tidal streams and be able to refine models of both intergalactic
enrichment and star formation triggering and regulation.

\acknowledgements %%% Text of acknowledgements runs on after this This
research was supported by Spitzer/NASA grants RSA No.s 1346930 (Higdon
\& Higdon), 1353814 (Smith) \& 1347980 (Struck).

\end{document}